\documentclass[a4paper]{jpconf}
\usepackage{graphicx}
\def\chandra{{\sl Chandra}}

\def\lesssim{\hbox{\rlap{\raise 0.425ex\hbox{$<$}}\lower
0.65ex\hbox{$\sim$} }}
\def\ltorder{\hbox{\rlap{\raise 0.425ex\hbox{$<$}}\lower
0.65ex\hbox{$\sim$}}}
\def\gtrsim{\hbox{\rlap{\raise 0.425ex\hbox{$>$}}\lower
0.65ex\hbox{$\sim$} }}

\def\xs{Arches}
\def\xq{Quintuplet}

\begin{document}

\title{Chandra Observations of Galactic Center: 
High Energy Processes at Arcsecond Resolution}

\author{Q. Daniel Wang$^1$}

\address{$^1$ Department of Astronomy, University of Massachusetts, Amherst, MA~01003}

\ead{wqd@astro.umass.edu}

\begin{abstract}
About 2 million seconds of \chandra\ observing time have been devoted to
the Galactic center (GC), including large-scale surveys and
deep pointings. These observations have led to the detection of 
about 4000 discrete X-ray sources and the mapping of diffuse X-ray emission
in various energy bands. In this review, I first summarize general 
results from recent studies and then present close-up views of the
three massive star clusters (Arches, Quintuplet, and GC) and their interplay 
with the Galactic nuclear environment.
\end{abstract}

\section{Introduction}

While Sgr A* itself is only weakly active at present, much of the 
high-energy activity in the GC is initiated apparently by 
the three young massive stellar clusters, located  within 50 pc radius 
of the super-massive black hole Sgr A* (Fig.~\ref{f:global}): Arches [with an age
of $(2-3) \times 10^6$ yrs], Quintuplet [$(3-6) \times 10^6$ yrs], and GC 
[$(3-7) \times 10^6$ yrs]
(e.g., Figer et al. 2004; Stolte et al. 2002, 2005; Genzel et al. 2003).
Massive stars themselves can be moderately bright X-ray sources 
(e.g., colliding
stellar wind binaries). Such stars also release large amounts of mechanical 
energy in form of fast stellar winds and supernovae, heating and shaping 
the surrounding interstellar medium (ISM) and affecting the accretion of 
the black hole. Furthermore, stellar end-products of massive stars (neutron stars and 
black holes) can also be strong X-ray sources. 
X-ray observations are thus a powerful tool for probing
such high-energy phenomena and processes.

The GC is a region where the spatial resolution of X-ray observations 
matters. This is why \chandra\ has invested heavily on the GC, 
conducting repeated 
large-scale raster surveys [existing 360 ks (Wang et al. 2002a) and upcoming
600 ks (PI: Muno)] 
and numerous deep pointed observations: about 1 Ms on Sgr A 
(Baganoff et al. 
2003), 100 ks each on Sgr B (Takagi et al. 2002), Sgr C (PI: Murakami), 
and Arches (Wang et al. 2006a), as well as 50 ks on Radio Arc (Yusef-Zadeh et al. 
2002a). The main instrument used in these \chandra\ observations is the 
ACIS-I, which covers a field of $17^\prime \times 17^\prime$ and
an energy range of 0.5-10 keV. But  below $\sim 2$ keV,
X-rays from the GC are heavily absorbed
by the ISM. The spatial resolution ranges 
from $\sim 1^{\prime\prime}$ on-axis to
$\sim 10^{\prime\prime}$ at the outer boundaries of the field.

While much of the data analysis is still ongoing, here I will first 
summarize some of the general results on detected discrete sources and 
diffuse X-ray emission and will then focus on 
the three massive star clusters, highlighting
various high-energy phenomena and processes involved.
 
\section{General results on point-like X-ray sources}

The \chandra\ observations have led to the detection of about 4000 
discrete sources; about 2400 of them are from the deep Sgr A 
observations (Wang et al. 2002a; Muno et al. 2003, 2006; Wang et al. 2006a).
Most of these sources are located physically in the vicinity of the GC, 
judged from their X-ray spectral characteristics and number statistics.
The number-flux relation (or the so-called log$N$-log$S$ relation) 
of the GC sources can be approximated as 
a power law with an index of $1.5\pm0.1$. But the relation is flatter 
(over the luminosity range of $5 \times 10^{31} {\rm~ergs~s^{-1}}\lesssim  L_x \lesssim 10^{34}
{\rm~ergs~s^{-1}}$) in the vicinity of 
the \xs\ and \xq\ clusters (Muno et al. 2006; Wang et al.
2006a), apparently due to the concentrations of massive stars and
possibly their end-products.

The GC X-ray sources represent a heterogeneous population of high-energy
objects. Bright sources (1E 1740.7-2942 and 1E 1743.1-2843) 
with 2-10 keV luminosities $L_x \gtrsim 10^{36}
{\rm~ergs~s^{-1}}$ are low-mass X-ray binaries (LMXBs). There is a clear
dearth of X-ray sources with 
$10^{34} {\rm~ergs~s^{-1}}\lesssim  L_x \lesssim 10^{36}
{\rm~ergs~s^{-1}}$, although a few transients are detected in 
this intermediate luminosity range  (e.g., Muno et al. 
2005; Sakano et al. 2005; 
Wijnands et al.  2006; Wang et al. 2006a). Sources within the range of 
$10^{33} {\rm~ergs~s^{-1}}\lesssim  L_x \lesssim 10^{34}
{\rm~ergs~s^{-1}}$ tend to be colliding wind massive
star binaries (Wang et al. 2006a) and possibly young pulsars (Wang et al. 2002b; Wang et al. 2006b). The nature of 
fainter sources ($10^{31} {\rm~ergs~s^{-1}}\lesssim  L_x \lesssim 10^{33}
{\rm~ergs~s^{-1}}$) 
are less certain; many of them are likely to be CVs, consistent with
the lack of  bright stellar counterparts in near-IR and radio 
surveys (e.g., Laycock et al.  2005; Bandyopadhyay 2005). A fraction of these sources have 
very hard X-ray spectra (e.g., with power law photon indices 
less than 0; Muno et al. 2004a). Such sources are most likely intermediate
polars, in which X-ray-emitting regions are at least partially obscured
and the observed X-rays represent the reflected emission
(Ruiter, Belczynski, \& Harrison 2006). 

\section{General results on diffuse X-ray emission}

Despite of their large number, the detected discrete X-ray sources typically 
account for only about 10\% of the total observed X-ray emission from the GC. 
The spectra of the remaining ``diffuse'' emission 
and the accumulated spectrum of the sources are similar, in terms of both
the continuum shape and the presence of prominent emission lines 
such as He-like S and Fe K$\alpha$ transitions, which 
indicate a broad thermal plasma temperature range of $\sim 1 - 10$ keV 
(Wang et al. 2002a; Muno et al. 2004b).
The spatial distribution of the diffuse emission appears to follow
closely the K-band stellar light (Muno et al. 2006). Therefore, a bulk of
the emission likely originates in the old stellar population (coronally
active binaries and CVs), 
as has been proposed for the Galactic ridge hard X-ray
emission (Revnivtsev et al. 2005).

There is no clear evidence for a significant presence of 
truly diffuse gas with $T \gtrsim 8$ keV, as was believed 
for many years. But a non-thermal diffuse hard X-ray component does seem
to be important (Wang et al. 2002a; see also the presentation by B. Warwick). 
The strongest evidence for this component is the ubiquitous
presence of the 6.4-keV line emission, globally correlated with the 
molecular gas in the GC region.
The line emission arises from the filling of inner shell vacancies 
of neutral or weakly ionized irons, which can be effectively  produced
by ionizing radiation with energies $\gtrsim 7.1$ keV. 
Because of the high equivalent width ($\gtrsim 1$ keV) of the observed 
line, the source of the ionizing radiation must be currently obscured or 
absent. This reflection interpretation gives 
the most convincing interpretation for the 6.4-keV line emission associated 
with the giant molecular clouds, Sgr B2 and Sgr C. 
The required ionizing source is believed to be
Sgr A*, which might be substantially brighter ($\gtrsim 10^{39} 
{\rm~ergs~s^{-1}}$) several hundred years ago. 
Indeed, the continuum emission from Sgr B, presumably due to 
the reflection through electron Thompson scattering, is shown to have a 
hard X-ray spectrum (power law photon index $\Gamma \approx 1.8$) in the 2-200 keV range,
consistent with what is expected from a typical AGN (Revnivtsev et al. 2004). 
However, the reflection 
interpretation is less
successful in explaining the 6.4-keV line emission observed in fields closer
to Sgr A* (e.g., Wang et al. 2002a). 
While the line intensity is globally correlated with the emission
from trace molecules, a peak-to-peak correlation is often absent. The lack of
such a correlation is not expected, because molecular clouds should be 
optically thin to the X-ray radiation. But it is possible that their density 
peaks may be optically thick to the emission
such as CS (J=2-1). 

An alternative mechanism to produce the
K-shell vacancies is the collision of irons with low 
energy cosmic ray electrons (LECRe; Valinia et al. 2000). In this case, one also expect a hard X-ray  emission from the bremsstrahlung process. 
The problem with this mechanism is the low efficiency of the emission
(typically $10^{-4} -  10^{-5}$); most of the energy
is consumed in the ionization loss. It is
not clear how the required energy density of the LECRe could be maintained 
{\sl globally} even in the GC region.

\section{\chandra\ view of the massive star clusters in the GC region}\label{ss:aqc}

\begin{figure}[ht]
\vspace{-4pc}
\includegraphics[width=36pc,angle=0]{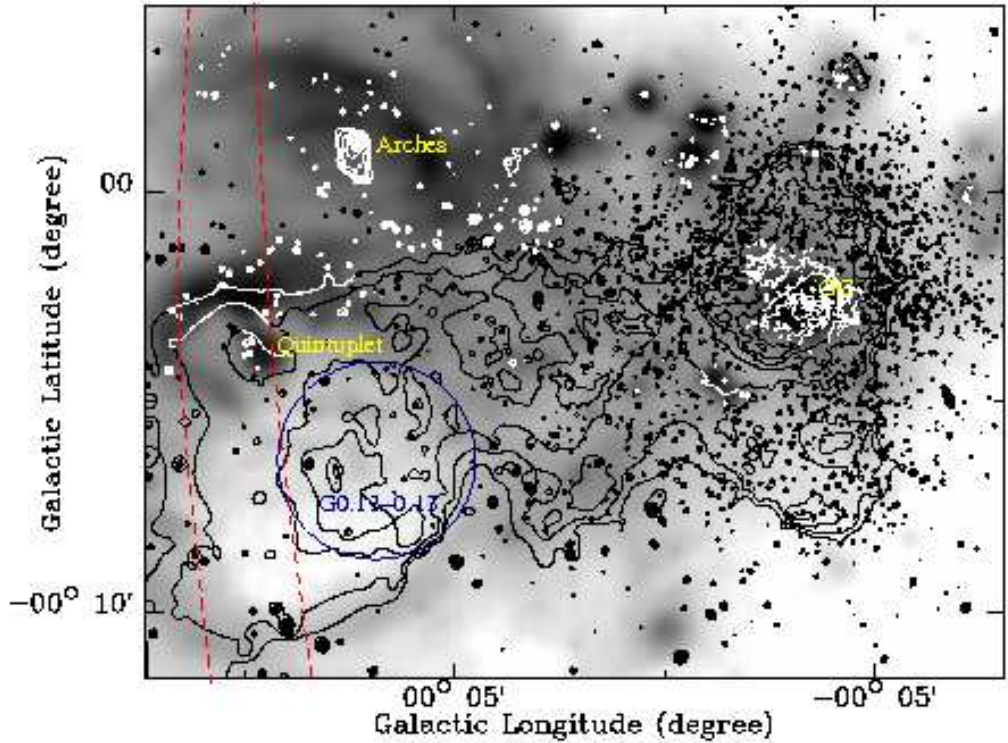}
\hspace{0pc}
\begin{minipage}[b]{38pc}
\caption{Panoramic views of the GC environment  of the three massive star clusters: 
MSX 24 $\mu$m intensity distributions (gray-scale; Price et al. 2001) 
and \chandra\ ACIS-I 1-9 keV intensity contours. The region with strong
nonthermal radio filaments are outlined by the dashed lines, while
the cloud G0.13-0.13 is circled.} \label{f:global}
\end{minipage}
\end{figure}

\begin{figure}[ht]
\hspace{6pc}\includegraphics[width=22pc,angle=0]{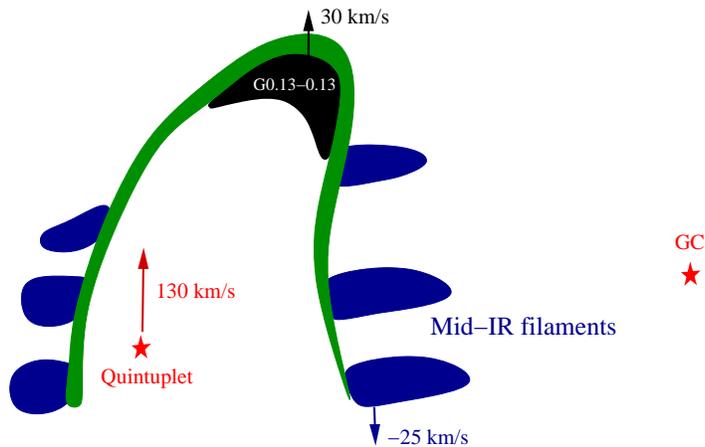}
\hspace{0pc}
\begin{minipage}[b]{38pc}
\caption{A plausible configuration of several
major ISM components in the vicinity of the \xq\ 
clusters, as viewed from the Galactic north pole. The
observed line-of-sight velocities of the components are marked.}
\label{f:ill_cloud}
\end{minipage}
\end{figure}

Fig.~\ref{f:global} presents the global perspective of the \xs\ and \xq\ clusters
 in the GC environment. This most
active region also includes various prominent thermal and 
nonthermal radio
filaments, dense molecular clouds, and strong diffuse X-ray emission, 
all of which are 
concentrated on the positive Galactic longitude side of Sgr A$^*$ (Wang et al.
2006a and references therein).  
This lopsided distribution of these features may partly be a 
chance coincidence. But some of 
the features are likely to be related, although they seem to have 
very different line-of-sight velocities. 
The thermal mid-IR-emitting filaments are clearly due
to the radiative heating of the \xs\ and \xq\ clusters, while the
strong 6.4-keV line emission is associated with the dense molecular
gas (Wang et al. 2002a; Yusef-Zadeh et al. 2002b). 
Fig.~\ref{f:ill_cloud}
illustrates a scenario that provides a unified interpretation of various
distinct interstellar features observed in the region. The densest cloud 
G0.13-0.13 in the region (Handa et al. 2006) is projected inside 
the west part of the distinct mid-IR cavity (Fig.~\ref{f:global};
Price et al. 2001) and is probably
located at the far side of a tunnel dug out by the cloud 
(Fig.~\ref{f:ill_cloud}).  This tunnel in projection is seen as the 
cavity  and
is apparently filled with hot plasma, responsible for 
the enhanced diffuse X-ray emission (Fig.~\ref{f:global}; Wang et al. 2006a).
This scenario is consistent with the lack of an enhanced near-IR extinction 
toward G0.13-0.13 and with the
high excitation of molecular gas associated with the cloud, apparently due to 
shock-heating (Handa et al. 2006).
The compression of the inter-cloud medium (the cavity wall in Fig.~\ref{f:ill_cloud}),
hence the attached magnetic field, may even explain the nonthermal radio
filaments. The diffuse X-ray emission, both thermal and nonthermal, are likely
products of the mechanical energy input from the massive star clusters. 
Much work is still  required to clarify what is actually going on
in this unusually dynamic region of the GC.
 
\subsection{The Arches and \xq\ Clusters}

\begin{figure}[ht]
\vspace{-2pc}
\includegraphics[width=38pc]{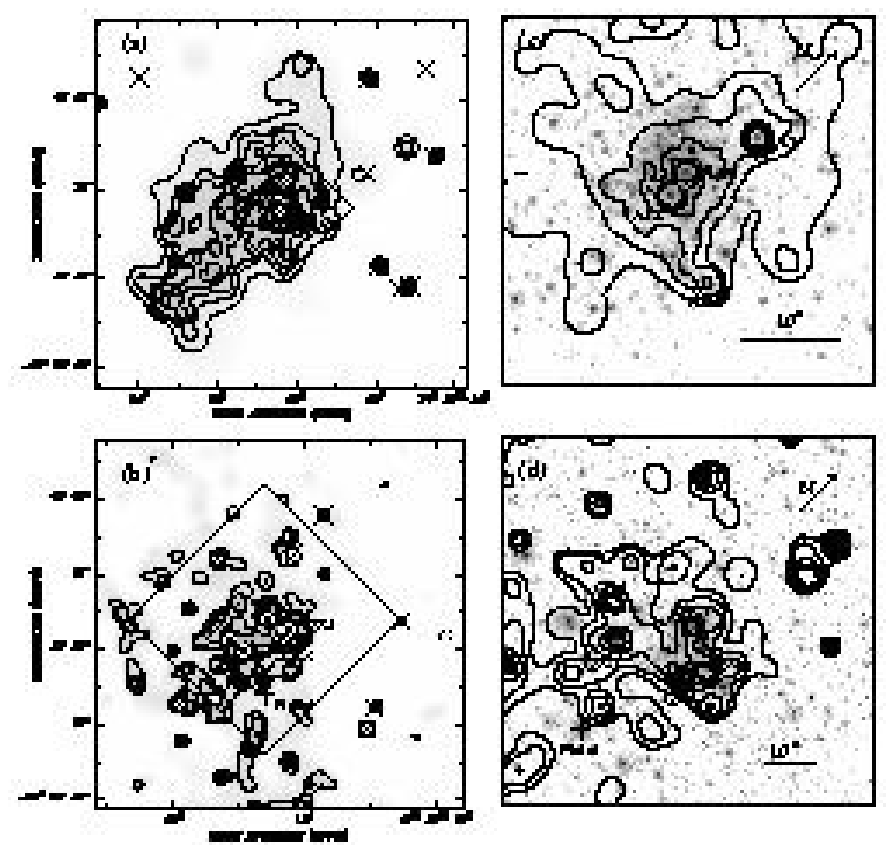}
\hspace{2pc}
\begin{minipage}[b]{38pc}
\caption{Adaptively smoothed ACIS-I 1-9 keV band images of the \xs\ (a) and
\xq\ (b) clusters. The intensity contour levels
are at 20, 23, 27, 33, 57, 57, 80, 114, 180,
314, 482, 682, 1351, and 3358 (above a local background of 13.4)
for (a), and at 17, 29, 33, 42, 54, and 72 (above 17) for
(b); all in units of $10^{-3} {\rm~counts~s^{-1}~arcmin^{-2}}$.
The two large squares in (a) and (b) outline the fields covered by
the {\sl HST} NICMOS near-IR images of the \xs\ (c) and \xq\ (d),
respectively (Figer et~al. 2004). The contours are the same as in (a) and (b),
except for excluding the first four levels in (c) for clarity.
The detected sources are marked with {\sl crosses} in (a) and (b). 
Several bright X-ray sources named previously are labeled 
(Yusef-Zadeh et al. 2002a; Law \& Yusef-Zadeh 2004).
}
\label{f:im_multi_bw}
\end{minipage}
\end{figure}

Both the \xs\ and \xq\ clusters are associated with local enhancements of 
X-ray emission (e.g., Fig.~\ref{f:im_multi_bw}), which was discovered
serendipitously at large off-axis angles in early 
{\sl Chandra} observations (Yusef-Zadeh et al. 2002a; Wang et al. 2002a;
Law \& Yusef-Zadeh 2004). A recent
100 ks observation allows for an in-depth study of both discrete and
diffuse components of the X-ray emission from these two clusters 
(Wang et al. 2006a).

\begin{figure}[ht]
\begin{minipage}{18pc}
\includegraphics[width=12pc,angle=270]{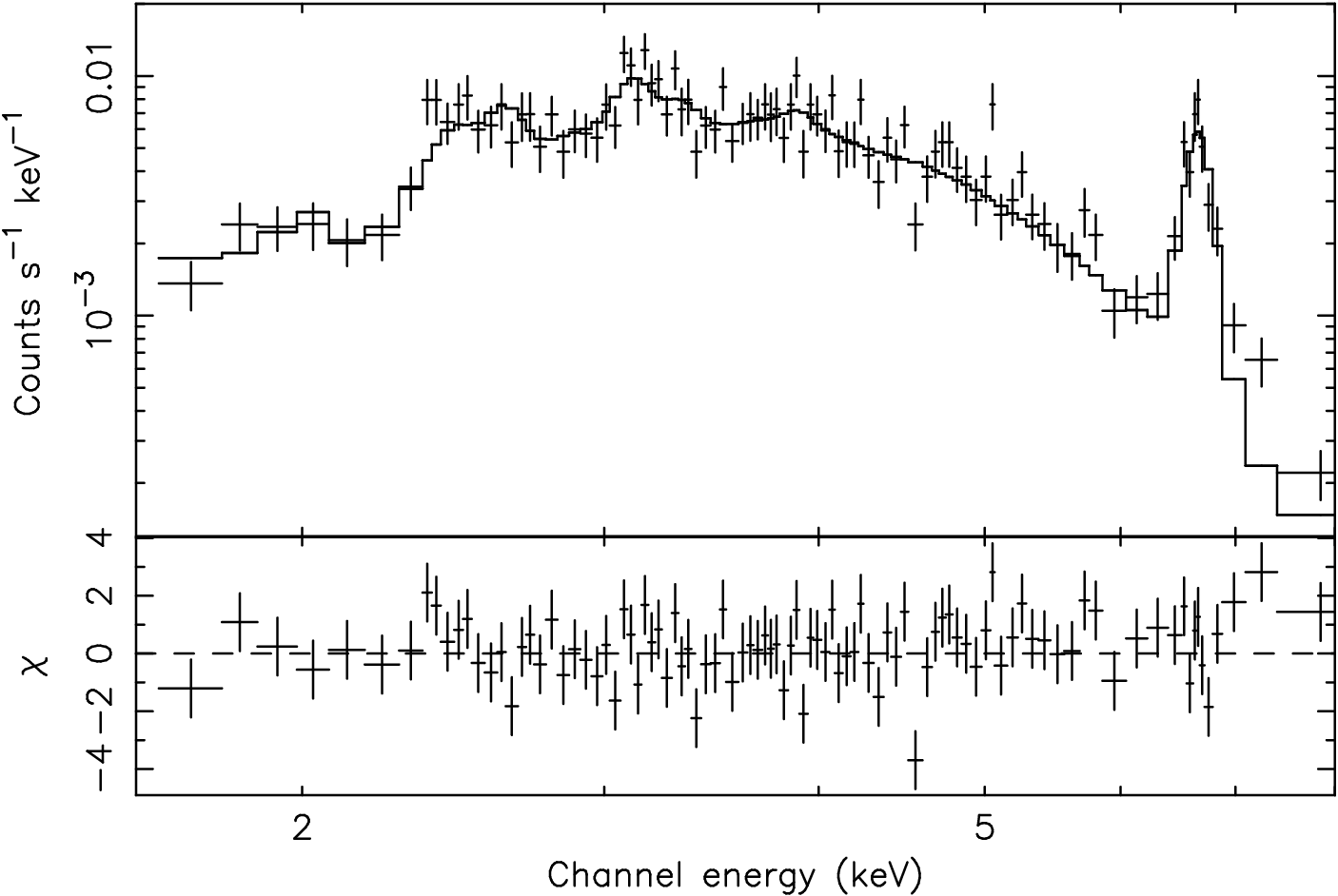}
\caption{Accumulated ACIS-I spectrum of the three brightest X-ray sources 
(A1N + A1S + A2) in the Arches cluster and the best-fit thermal plasma 
models.} \label{f:spec_a}
\end{minipage}
\hspace{1pc}
\begin{minipage}{18pc}
\includegraphics[width=12pc,angle=270]{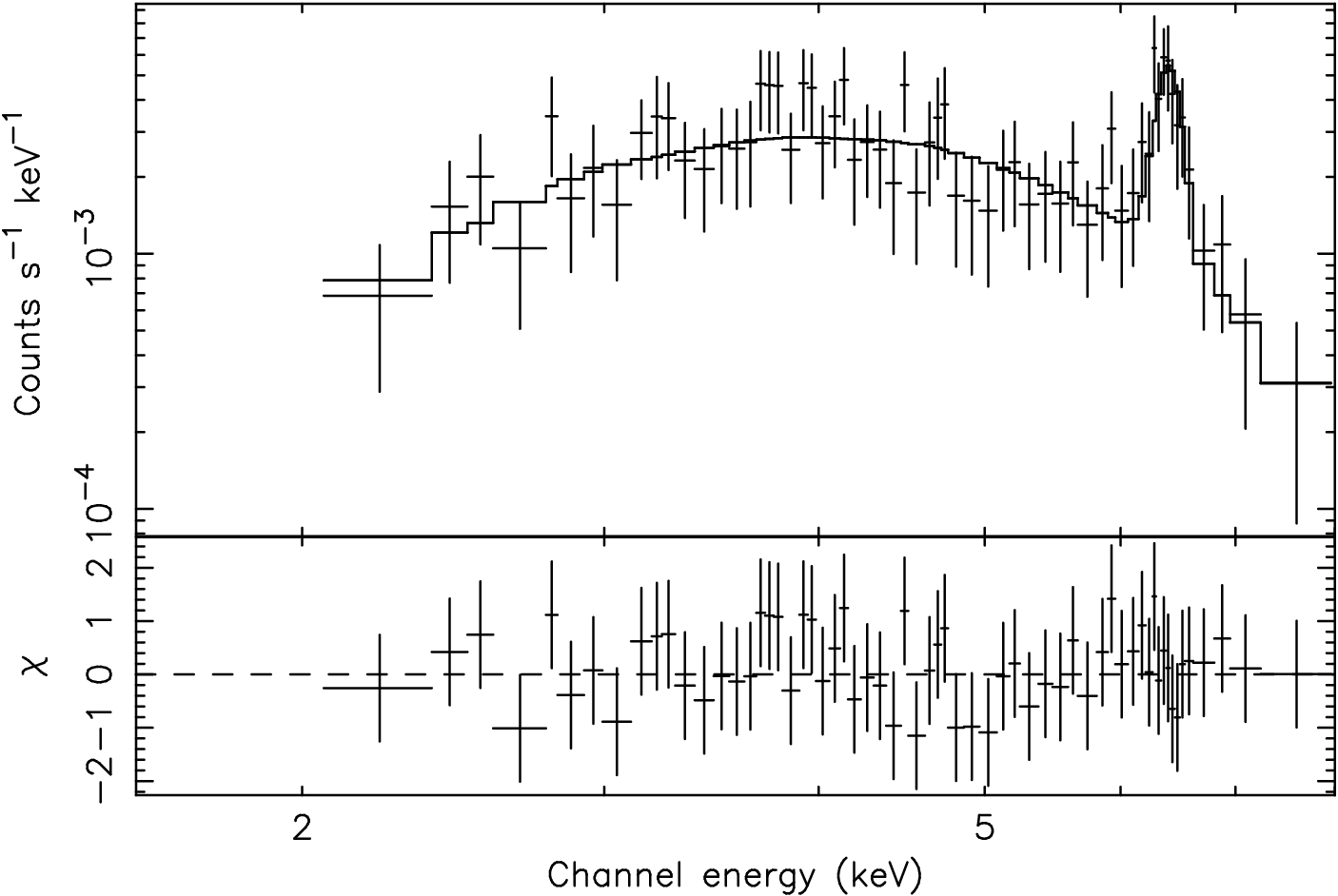}
\caption{ ACIS-I spectrum of the diffuse X-ray emission southeast 
of the Arches cluster (Fig.~\ref{f:cs_dx})
and the best-fit power law plus 
6.4-keV Gaussian line model.}
\label{f:spec_aline}
\end{minipage} 
\end{figure}

The three bright X-ray sources in the \xs\ core region all have near-IR 
counterparts, classified as WN stars (Fig.~\ref{f:im_multi_bw}b). 
These sources have remarkably
similar spectra, in terms of both the continuum shape and the strong
presence of the 6.7-keV line (Fig.~\ref{f:spec_a}). The spectrum can be 
characterized by
an optically-thin thermal plasma with a temperature of $\sim 2$ keV and 
a metal (chiefly iron) abundance of 1.8$\times$ solar. The 0.3-8 keV luminosity
of each source is $\sim 1 \times 10^{34} {\rm~ergs~s^{-1}}$. With these 
properties, the sources are most likely colliding wind massive star
binaries, although their luminosities are somewhat higher than all known 
such objects. The X-ray metal abundance measurement, based primarily on
the He-like Fe K$\alpha$ line,
is also interesting, which is insensitive to
the exact temperature of the plasma. The optically-thin 
thermal emission process
is also quite simple, astrophysically. Furthermore, 
the iron abundance in the stellar winds
should not be contaminated by the nuclear synthesis of these stars and
thus reflect the value in the ISM of the GC. 

\begin{figure}[ht]
\vspace{-2pc}
\begin{minipage}{18pc}
\includegraphics[width=18pc]{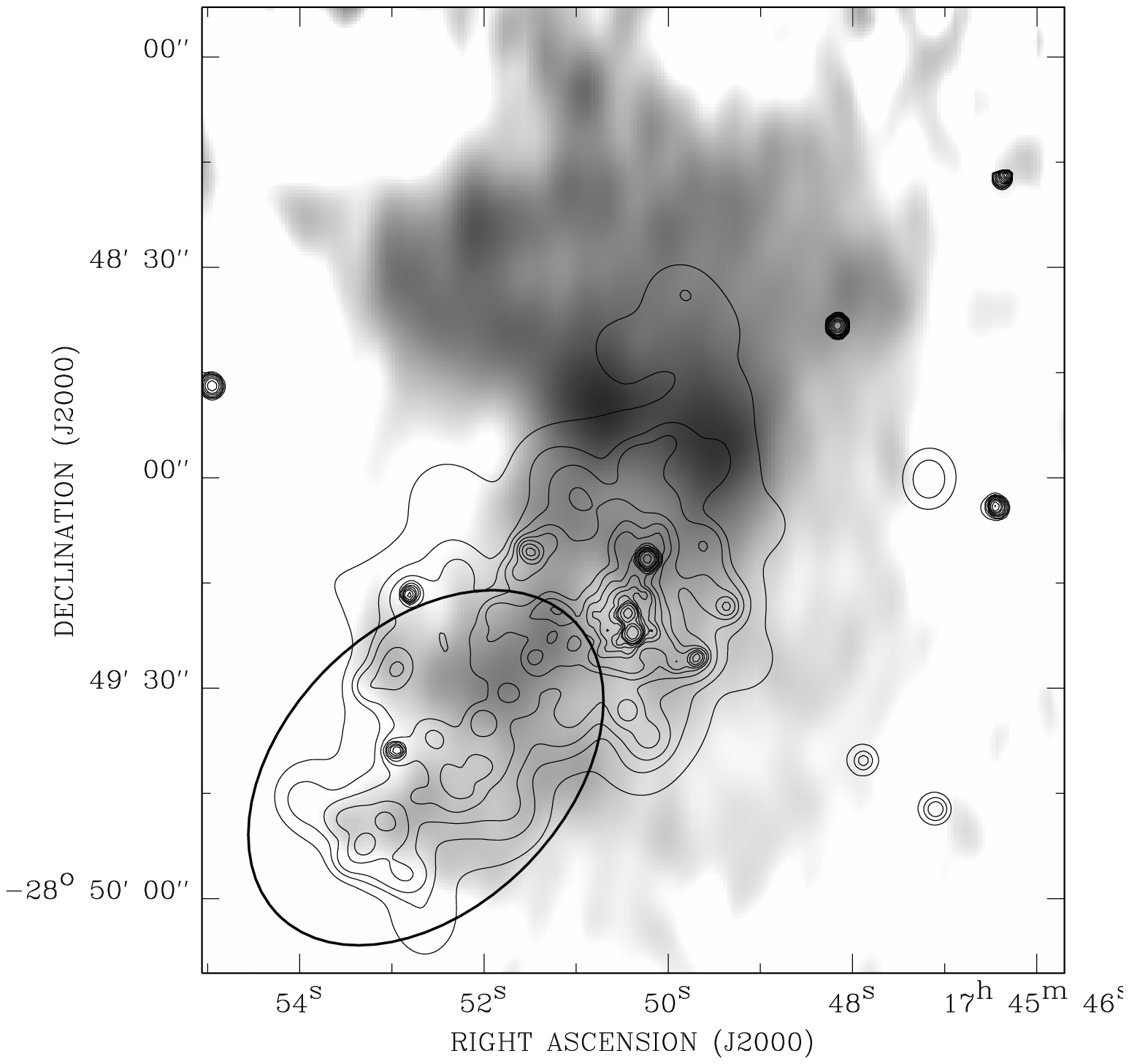}
\caption{\label{f:cs_dx}ACIS 1-9 keV diffuse emission intensity contours overlaid on the CS (J=2-1) line emission image. The X-ray contours are the same as
in Fig.~\ref{f:im_multi_bw}a. The ellipse outlines
the region from which the spectrum in Fig.~\ref{f:spec_aline} is
extracted.}
\vspace{-2pc}
\end{minipage}
\hspace{2pc}
\vspace{3pc}
\begin{minipage}{18pc}
\hspace{2pc}
\includegraphics[width=11pc]{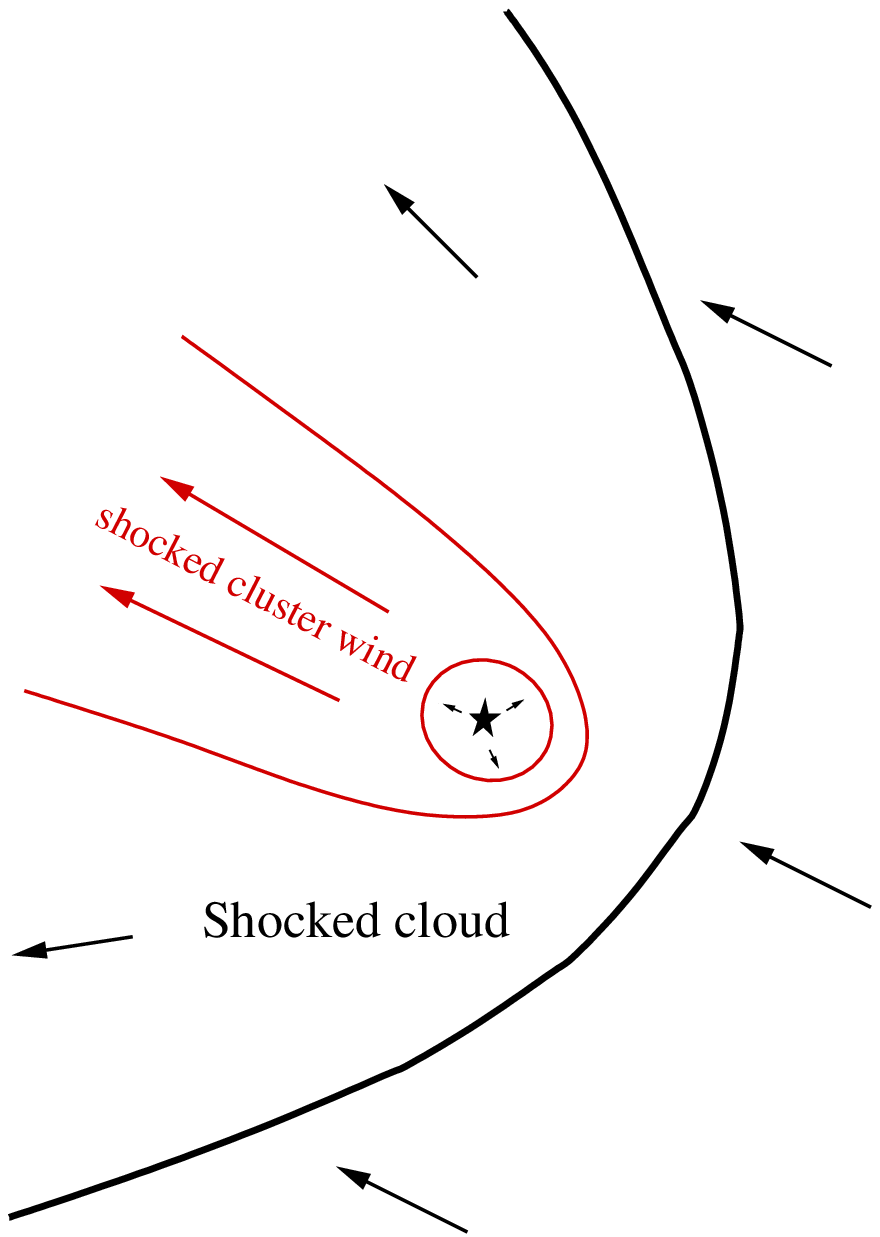}
\vspace{2pc}
\caption{\label{f:ill} An illustration of the proposed cluster-cloud 
collision scenario. The shocked cloud gas is partly traced by the CS 
and 6.4-keV lines (Fig.~\ref{f:cs_dx}), whereas the shocked cluster
wind plasma near the cluster is by the 6.7-keV line. }
\vspace{-4pc}
\end{minipage} 
\end{figure}

\begin{figure}[!htb]
\includegraphics[width=38pc]{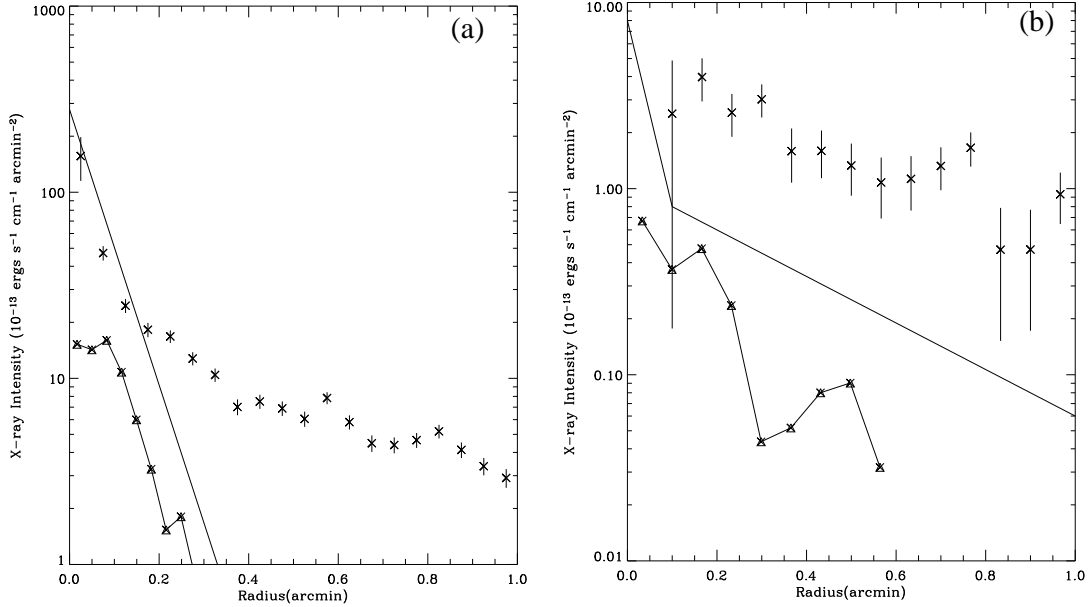}
\begin{minipage}[b]{38pc}  
\caption{\label{f:rbp} Radial ACIS-I 1-9 keV intensity profiles ({\sl crosses} with $1\sigma$ error
bars) around the \xs\ (a) and \xq\ (b) clusters, compared with the
respective NICMOS F205W stellar light distributions (connected
{\sl triangles}). The cluster wind predictions are shown
approximately as the solid line
from 3-D simulations for the ``standard'' stellar wind mass-loss rates of the
two clusters (Rockefeller et al. 2005)}.
\vspace{-2pc}
\end{minipage}
\end{figure}

The nature of the source-removed ``diffuse'' emission is more complicated. The 
overall spectrum of the emission shows both
6.4-keV and 6.7-keV emission lines. The 6.7-keV line arises predominately
in the region close to the core of the \xs\ cluster. The
region has an extent of $\sim 30^{\prime\prime}$ and appears to be 
elongated towards the east, morphologically matching 
an extinction deficit around the cluster 
(Stolte et al. 2002). The 6.4-keV line emission is more widely distributed,
but is particularly enhanced in the southeast extension of
the diffuse X-ray emission (Fig.~\ref{f:im_multi_bw}a and 
Fig.~\ref{f:spec_aline}). All these can be interpreted as the collision 
between the cluster wind of the \xs\ cluster and a
dense gas cloud (Fig.~\ref{f:ill}). The steep decline of the observed 
surface intensity of the diffuse X-ray emission with the off-cluster 
radius within $\sim 10^{\prime\prime}$ is consistent with
the prediction of the emission from the expanding cluster wind
(Fig.~\ref{f:rbp}a), whereas
the flattening of the intensity distribution at larger radii 
(but $\lesssim 15^{\prime\prime}$) likely reflects
the reverse shock heating and confinement of the wind.

The presence of the cloud in the vicinity of the \xs\ cluster is known. 
In the CS map made by Serabyn \& Gusten (1987) with the 30 {\sl IRAM} telescope, the cloud is 
labeled as ``Peak 2''. Fig.~\ref{f:cs_dx} shows a recent interferometry map made with 
{\sl OVRO} (Wang et al. 2006a). The cloud has a line-of-sight velocity of $-25 {\rm~km~s^{-1}}$ 
(i.e., moving toward us), which is similar to many other clouds or 
filaments in the region. In comparison, the cluster is moving at $\sim 95 
{\rm~km~s^{-1}}$ away from us. Therefore, the relative velocity between
the two is at least about $120 {\rm~km~s^{-1}}$. Because the filling 
factor of molecular clouds in the region is estimated to be about 0.3 
(Serabyn \& Gusten 1987),
a chance collision between the cluster and such a cloud is not rare.
The far-IR spectroscopy further shows the presence of
a component of dusty gas at a velocity of $-70 
{\rm~km~s^{-1}}$, unique at the location of the \xs\ cluster
(Cotera et al. 2005). This component
may represent shocked cloud gas, deflected toward us. 

The collision between the cluster and the cloud (Fig.~\ref{f:ill}) also provides a plausible 
explanation of the large-scale diffuse X-ray emission (Fig.~\ref{f:cs_dx}).  Indeed, the
overall morphology of the diffuse X-ray emission resembles a bow
shock. Presumably, the motion of the cloud relative to the cluster
is roughly toward the east in the sky. The diffuse X-ray emission, except in
the core region around the \xs\ cluster,  is reasonably well correlated 
with the 6.4-keV line intensity,  but shows little peak-to-peak 
correlation with the CS line intensity. The diffuse X-ray emission is 
enhanced particularly in the region southeast of the cluster,
probably tracing the eastern edge of the cloud. 
But the northern part of the cloud, where the CS line emission is the strongest,
shows no enhancement in the 6.4-keV line emission. The cloud does not 
seem to be dense enough to be optically thick to the CS line emission. Therefore,
the radiation reflection interpretation does not work in this case. Certainly,
the source of the radiation cannot be the \xs\ cluster itself. The luminosity 
of the cluster falls several orders of magnitude
short of the required. The most probable origin of the diffuse X-ray 
X-ray emission is then LECRe that may be produced locally in and around
the clusters (Wang et al. 2006a). 

The older and looser \xq\ cluster is weak in X-ray emission, 
both point-like and diffuse (Fig.~\ref{f:im_multi_bw}b and d, Fig.~\ref{f:rbp}b).
The X-ray sources in the core of the cluster 
also show more diverse spectral
characteristics, with substantially different intrinsic spectral shapes
and absorptions. Some of these sources are probably colliding 
dusty winds in massive star binaries. The Pistol star, despite
of its enormous bolometric luminosity, is not detected. But the 
3$\sigma$ upper limit to the 0.3-8 keV luminosity, 3$\times10^{33}$
${\rm~ergs~s^{-1}}$, is consistent with the nominal relation
$L_{x}/L_{bol} \sim 10^{-7}$ for massive stars.
The diffuse X-ray emission from the \xq\ cluster ($L_X \sim 
2 \times 10^{33} {\rm~ergs~s^{-1}}$) is about a factor of 10 lower than that
from the \xs\ cluster and can be naturally explained by the cluster wind and
a limited number of low-mass pre-main sequence young stellar objects (YSOs).

There appears to be a general deficiency of YSOs in the two
clusters, relative to the prediction from 
the standard Miller \& Scalo initial mass function (IMF). 
Compared with the X-ray emission from 
YSOs in the Orion nebula (Feigelson et al. 2005), 
the observed total diffuse X-ray luminosities from
the Arches and Quintuplet clusters suggest that they contain no
more than $2 \times 10^4$ and  $3 \times 10^3$ YSOs. These numbers are 
a factor of 10 and 5 smaller than what would be expected from the
IMF and the massive star populations observed in the cores of the two clusters. 
One possibility is that the IMF flattens at intermediate masses,
as indicated in a near-IR study of the inner regions of the \xs\ 
cluster (Stolte et al. 2005).  The top heavy IMF may be a result of
the star formation in the extreme environment of the GC. In particular,
collisions between dense molecular clouds such as G0.13-0.13 
may be responsible for the formation of massive 
clusters. 

\subsection{The GC cluster}

\begin{figure}[ht]
\begin{minipage}{14pc}
\includegraphics[width=14pc,angle=0]{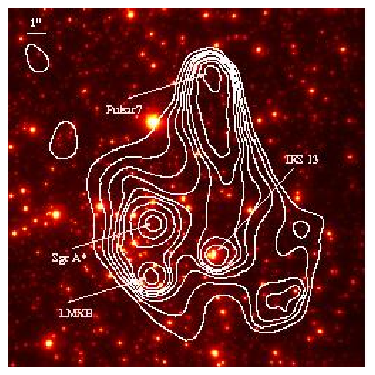}
\caption{\label{f:gc_x} SINFONI near-IR image of the GC cluster
(Eisenhauer et al. 2005) with overlaid ACIS-I 1--9~keV  intensity contours 
 (Wang et al. 2006b).}
\end{minipage}
\hspace{1pc}
\begin{minipage}{22pc}
\includegraphics[width=22pc,angle=0]{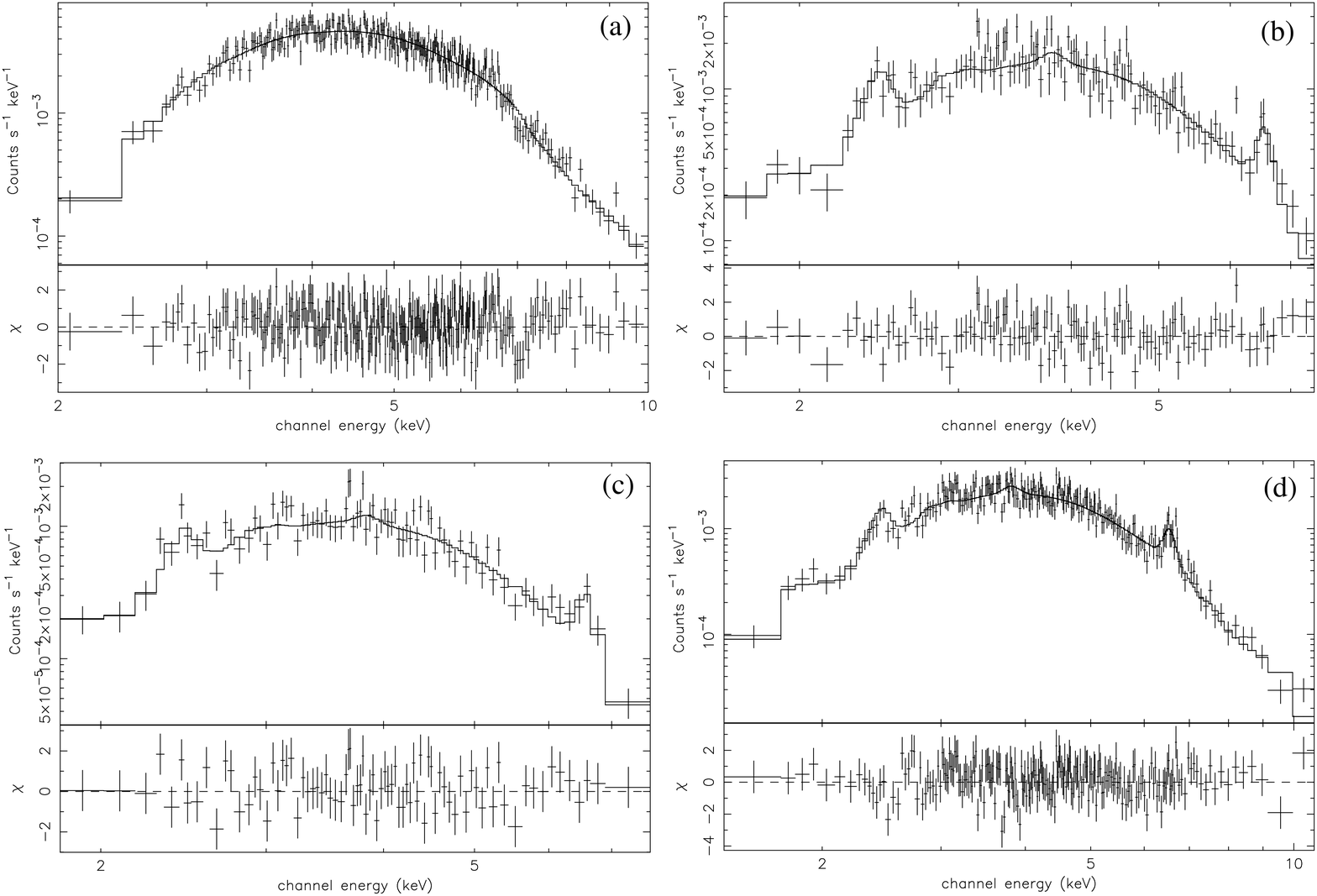}
\caption{ACIS-I spectra of \xs\ (a),
Diffuse (b), IRA 13 (c), and Sgr A$^*$ (d). The relative deviations 
from the best-fit models (Wang et al. 2006b) are shown in the 
respective bottom panels.
\label{f:spec}}
\end{minipage} 
\end{figure}

The deficiency of YSOs in the GC cluster has been 
proposed by Nayakshin et al. (2005), based on the 
observed diffuse X-ray intensity of the region (Fig.~\ref{f:gc_x};
Baganoff et al. 2003). 
From the X-ray constraint on the total mass of the cluster, 
they further conclude that
it cannot be a remnant of a massive star cluster that originated
at several tens of parsecs away from Sgr A* and then
dynamically spiralled in. The GC cluster was thus most likely formed
{\sl in situ} in a self-gravitating circum-nuclear disk.
Therefore, the top heavy IMF appears to be a general characteristic of
star formation in the GC region. But the exact form
of the IMF is yet to be determined for the clusters.

The diffuse X-ray emission from the GC cluster region appears significantly different
from those from the \xs\ and \xq\ clusters. The spectrum of the 
emission shows a Fe K$\alpha$ emission line at an intermediate
energy $\sim 6.55$ keV (e.g., Fig.~\ref{f:spec}b; Baganoff et al. 2003), which is a clear
signature for the non-equilibrium ionization
(NEI) state of the X-ray-emitting plasma with an ionization timescale
$n_e t \sim 3 \times 10^{10} {\rm~cm^{-3}~s}$  (Wang et al. 2006b).
The source of the plasma is likely to be the stellar winds from 
massive stars in the GC cluster. We may assume that the pre-shock stellar winds are
cold ($T \lesssim 10^4$ K) and generally have velocities of
$\sim 1 \times 10^3 {\rm~km~s^{-1}}$). Fig.~\ref{f:gc_x} shows that the 
distribution of the stars is not 
very centrally concentrated, except for the compact
sub-cluster complex IRS 13. The density of the shocked wind plasma is typically not very high (order 
$\sim 10  {\rm~cm^{-3}}$). Thus, much of the shock-heated plasma may be out of 
the collisional ionization equilibrium, at least in individual 
wind-wind shock regions. 

Further constraints on the properties of the plasma may  be obtained from 
studying its interaction with individual stars or sub-clusters. The X-ray study
of IRS 13 provides such a possibility. The X-ray emission from this sub-cluster is
clearly resolved and shows
a non-NEI plasma spectrum (Fig.~\ref{f:spec}c). Therefore, the source most likely
represents stellar wind-wind collisions. The X-ray emission
may be enhanced due to a strong ram-pressure confinement by the
overall GC cluster wind, plus a potential outflow from Sgr A*, as 
predicted in the radiatively inefficient accretion flow models 
(Yuan, Quataert, \& Narayan 2003).
The confinement is indicated by the X-ray morphology that 
is slightly offset from the centroid position of IRS 13
and elongated toward to the west (Wang et al. 2006b). 

The properties of the plasma around Sgr A* should also affect 
its accretion. The quiescent X-ray emission from Sgr A* is
resolved to have a size of about a couple of arc-seconds. This size corresponds
to the Bondi accretion radius of the super-massive black hole. The accretion 
flow is believed to be responsible for the X-ray emission. 
Interestingly, the observed spectrum of the emission resembles that of the 
surrounding diffuse plasma
and shows the Fe K$\alpha$ line at $\sim 6.6$ keV (Fig.~\ref{f:spec}d; Wang et al. 2006b; 
Xu et al. 2006). Therefore, the flow is also in
an NEI state. A detailed modeling of the flow, confronted with
the observed spatial and spectral distributions of the X-ray emission,
will help to constrain the accretion dynamics.

The X-ray emission from the GC region is complicated by the
presence of a comet-like pulsar wind nebula (PWN), a few 
arc-seconds northwest of Sgr A* (Fig.~\ref{f:gc_x}; Wang et al. 2006b). 
The suspected pulsar, 
presumably moving at a high speed relative to the ambient medium, is located 
at the northern head of the nebula. This PWN interpretation naturally explains not only the morphology, 
but also the nonthermal spectrum (Fig.~\ref{f:spec}a) 
and the spectral steepening with the off-source distance due to the 
synchrotron cooling of ultra-relativistic electrons (and positrons). 
Furthermore, the inverse-Compton scattering of the intense ambient
infrared photon field by the electrons provides a ready explanation for the 
TeV emission from the GC (Aharonian et al.  2004). The lack of a 
synchrotron-emitting radio or infrared counterpart of the PWN is at 
least partly due to the fast Compton-cooling 
(hence little accumulation) of relatively low-energy electrons, for which 
the Klein-Nishina suppression of the inverse-Compton scattering efficiency
is not important. The presence of this young pulsar 
raises a number of questions: where does it originate? and where is the
supernova remnant? The most logic origin of the pulsar is the
GC cluster itself. But the remnant is more difficult to isolate, the properties
of which depend sensitively on the density and temperature structure of 
the medium (e.g., Tang \& Wang 2005). Considering the orbital 
motion around Sgr A*, it is even possible that the pulsar may be produced 
together with the well-known SNR Sgr A West, although an alternative pulsar candidate 
has already been proposed for the remnant (see the contribution by S. Park).

Clearly, the above is a very incomplete and subjective review of recent
{\sl Chandra} results on the GC. We will certainly 
learn a great deal more from the existing and upcoming {\sl Chandra} 
observations before GC-2009 Workshop in China.
 
\ack
Some of the ideas presented above
are developed during or after many stimulating discussions with workshop 
participants, to whom I am grateful.
I also thank my collaborators for their contributions to the
various research projects mentioned above and the organizers of the workshop
for the invitation and the hospitality. This work is supported by NASA through the grant SAO/CXC GO4-5010X.

\section*{References}
{\scriptsize
Aharonian, F., et al.  2004, A\&AL, 425, 13\\
Baganoff, F. K., et al. 2001, Nature, 413, 45\\
Baganoff, F. K., et al. 2003, ApJ, 591, 891\\
Bandyopadhyay, R. M, et al. 2005, MNRAS, 364, 1195\\
Cotera, A. S., Colgan, S. W. J., Simpson, J. P., \& Rubin, R. H. 2005, ApJ, 622, 333\\
Eisenhauer, F., et al. 2005, ApJ, 628, 246\\
Feigelson, E. D., et al. 2005, ApJS, 160, 379\\
Figer, D. F., Rich, R. M., Kim, S. S., Morris, M., \& Serabyn, E. 2004, ApJ, 601, 319\\
Genzel, R., et al. 2003, ApJ, 594, 812\\
Handa, T., et al. 2006, ApJ, 636, 261\\
Law, C., \& Yusef-Zadeh, F. 2004, ApJ, 611, 858\\
Laycock, S., et al.  2005, ApJL, 634, 53\\
Muno, M. P., et al. 2003, ApJ, 589, 225\\
Muno, M. P., et al.  2004a, ApJ, 613, 1179\\
Muno, M. P., et al.  2004b, ApJ, 613, 326\\
Muno, M. P., et al.  2005, ApJL, 622, 113\\
Muno, M., Bauer, F. E., Bandyopadhyay, R. M., \& Wang, Q. D. 2006, ApJ, in press (astro-ph/0601627)\\
Nayakshin, S., \& Sunyaev, R. 2005, MNRAS, 364, 23\\
Price, S. D., et~al.  2001, ApJ, 121, 2819\\
Revnivtsev, M. G., et al. 2004, A\&AL, 425, 49\\
Revnivtsev, M. G., et al. 2005, A\&A, submitted (astro-ph/0510050)\\
Rockefeller, G., Fryer, C. L., Melia F., \& Wang, Q., D. 2005, ApJ, 623, 171\\
Ruiter, A. J., Belczynski, K., \& Harrison, T. E. 2006, ApJL, 640, 167\\
Serabyn, E., \& G\"uesten, R. 1987, A\&A, 184, 133\\
Sakano, M., Warwick, R. S., Decourchelle, A., \& Wang, Q. D. 2005, MNRAS, 357, 1211\\
Stolte, A., Grebel, E. K., Brandner, W., \& Figer, D. F. 2002, A\&A, 394, 45\\
Stolte, A., Brandner, W., Grebel, E. K., Lenzen, R., \& Lagrange, A. 2005, ApJL, 628, 113\\
Takagi, S., Murakami, H., \& Koyama, K. 2002, ApJ, 573, 275\\
Tang, S. K., \& Wang, Q. D. 2005, ApJ, 628, 205\\
Valinia, A., Tatischeff, V., Arnaud, K., Ebisawa, K., \& Ramaty, R. 2000, ApJ, 543, 733\\
Wang, Q. D., Gotthelf, E. V., \& Lang, C., 2002, Nature, 415, 148\\
Wang, Q. D., Lu, F. J., \& Lang, C. C. 2002, ApJ, 581, 1148\\
Wang, Q. D., Dong, H., \& Lang, C. C. 2006a, MNRAS, in press (astro-ph/0606282)\\
Wang, Q. D., Lu, F. J., \& Gotthelf, E. V. 2006b, MNRAS, 367, 937\\
Wijnands, R., et al.  2006, A\&A, 449, 1117\\
Xu, Y. D., et al. 2006, ApJ, 640, 319\\
Yuan, F. Quataert, E., \& Narayan, R. 2003, ApJ, 598, 301\\
Yusef-Zadeh, F., et~al., 2002a, ApJ, 570, 665\\
Yusef-Zadeh, F., Law, C., \& Wardle, M. 2002b, ApJ, 568, 121
}
\end{document}